\documentstyle[twoside,fleqn,espcrc2,fps]{article}

\newcommand{\AmS}{{\protect\the\textfont2
  A\kern-.1667em\lower.5ex\hbox{M}\kern-.125emS}}

\def\lsi{\raise0.3ex\hbox{$<$\kern-0.75em\raise-1.1ex\hbox{$\sim$}}}
\def\gsi{\raise0.3ex\hbox{$>$\kern-0.75em\raise-1.1ex\hbox{$\sim$}}}

\newcommand{\R}{{\kern+.25em\sf{R}\kern-.78em\sf{I} 
  \kern+.78em\kern-.25em}}
\newcommand{\C}{{\kern+.25em\sf{C}\kern-.50em\sf{I} \kern+.50em\kern-.25em}}

\newcommand{\be}{\begin{equation}}
\newcommand{\ee}{\end{equation}}
\newcommand{\bea}{\begin{eqnarray}}
\newcommand{\eea}{\end{eqnarray}}
\newcommand{\nn}{\nonumber}
\newcommand{\bd}{\begin{displaymath}}
\newcommand{\ed}{\end{displaymath}}

\newcommand{\tgh}{{\rm tgh}}

\title{Naturally Light Fermions from Dimensional Reduction}

\author{W. Bietenholz
\address{ Institut f\"{u}r Physik, Humboldt Universit\"{a}t zu Berlin,
Newtonstr. 15, D-12489 Berlin, Germany \\
$^{{\rm b}}$ Institut f\"{u}r Theoretische Physik, Universit\"{a}t Bern, 
Sidlerstr. 5, CH-3012 Bern, Switzerland }
, A. Gfeller $^{{\rm b}}$ and U.-J. Wiese $^{{\rm b}}$
\thanks{Talk presented by W.B. at Lattice2003 \newline
\hspace*{1.2mm} Preprint HU-EP-03/48}
}

\begin{document}

\begin{abstract}

We consider the 3-d Gross-Neveu model in the broken phase
and construct a stable brane world by means of a domain
wall and an anti-wall. Fermions of opposite chirality
are localized on the walls and coupled through the 3-d bulk.
At large wall separation $\beta$ the 2-d correlation length
diverges exponentially, hence a 2-d observer cannot 
distinguish this situation from a 2-d space-time. The 3-d 4-fermion 
coupling and $\beta$ fix the effective 2-d coupling such
that the asymptotic freedom of the 2-d model arises. This mechanism
provides criticality without fine tuning.

\vspace*{-4mm}

\end{abstract}

\maketitle


\section{Motivation}

In Yang-Mills gauge theory on a lattice of spacing $a$,
one naturally obtains for the correlation length $\xi_g \gg a$
based on asymptotic freedom. However, if one includes quarks, say
as Wilson fermions, one would naturally arrive at $\xi_q = O(a)$, 
i.e.\ a small quark mass $\xi_q^{-1}$ requires an unnatural process of fine 
tuning. In this sense, Domain Wall Fermions \cite{DWF} approach the chiral
limit in a ``natural'' way, starting from a higher dimension.
However, there the ``extra dimension'' is
just technical: it does not host gauge fields, and one separates the
walls at fixed glueball mass $\xi_g^{-1}$.

Here we try to construct a brane world in an odd dimension, which is
endowed with a mechanism for dimensional reduction to
an even dimension with light fermions. This would represent an all 
natural mechanism for the hierarchy of scales.

In particular we are going to study the Gross-Neveu model at large $N$
and its reduction from $d=3$ to 2. Details
will be presented in Ref.\ \cite{BGW}.

\section{The 2-d Gross-Neveu model}

The Euclidean action of our target theory reads
\be  \label{act2d}
S [\bar \Psi , \Psi ] = \int d^{2}x \Big[ \bar \Psi \gamma_{\mu}
\partial_{\mu} \Psi - \frac{g}{2N} (\bar \Psi \Psi)^{2} \Big] \ ,
\ee
where we suppress the flavor index $1 \dots N$. It has a discrete
chiral $Z(2)$ symmetry
\bd
(\bar \Psi_{L}, \Psi_{L}) \to \pm (\bar \Psi_{L}, \Psi_{L}) , \,
(\bar \Psi_{R}, \Psi_{R}) \to \mp (\bar \Psi_{R}, \Psi_{R}) .
\ed
With an auxiliary scalar field $\Phi = \frac{G}{N} \bar \Psi \Psi$
the action (\ref{act2d}) is equivalent to
\bd
S [\bar \Psi , \Psi , \Phi ] = \int d^{2}x \Big[ \bar \Psi \gamma_{\mu}
\partial_{\mu} \Psi - \Phi \bar \Psi \Psi + \frac{N}{2g} \Phi^{2} \Big] 
\ .
\ed
In the limit $N \to \infty$, $\Phi$ freezes to a constant, and
$\bar \Psi , \Psi$ can be integrated out. This yields an effective
potential $V_{eff}(\Phi )$ with the minima $\pm \Phi_{0}$, which obey
the gap equation
\be
\frac{1}{g} = \frac{1}{\pi} \int_{0}^{\Lambda_{2}} dk \,
\frac{k}{k^{2} + \Phi_{0}^{2}} \ .
\ee
At weak coupling $g \ll 1$ we have $\Lambda_{2} \gg \Phi_{0}$ and
\be
m = \Phi_{0} = \Lambda_{2} e^{-\pi /g}
\ee
represents the  fermion mass, which is generated by the spontaneous
breaking of the $Z(2)$ symmetry. 
The exponent expresses asymptotic freedom.

\section{The 3-d Gross-Neveu model}

The 3-d action
\bd
S [\bar \Psi , \Psi ] \! = \!\! \int \!\!
d^{3}x \Big[ \bar \Psi \gamma_{\mu}
\partial_{\mu} \Psi + \bar \Psi \gamma_{3} \partial_{3} \Psi
- \frac{G}{2N} (\bar \Psi \Psi)^{2} \Big]
\ed
still has a $Z(2)$ symmetry,
\bea
(\bar \Psi_{L}, \Psi_{L})\vert_{(\vec x, x_{3})} & \to & 
\pm (\bar \Psi_{L}, \Psi_{L})\vert_{(\vec x, -x_{3})} \ ,
\nn \\  \label{Z2sym}
(\bar \Psi_{R}, \Psi_{R})\vert_{(\vec x, x_{3})} & \to &
\mp (\bar \Psi_{R}, \Psi_{R})\vert_{(\vec x, -x_{3})} \ ,
\eea
which turns into the discrete chiral symmetry 
after dimensional reduction.

The 3-d gap equation reads
\be  \label{3dgap}
\frac{1}{G} = \frac{1}{(2 \pi )^{3}} \int d^{3}k \, \frac{2}{k^{2}+
\Phi_{0}^{2}} \ ,
\ee
and for a cut-off $\Lambda_{3} \gg \Phi_{0}$ one identifies a critical
coupling $G_{c} = \pi^{2}/\Lambda_{3}$. At $G > G_{c}$ we are in the
broken phase with 
\be
\Phi_{0} = 2\pi ( 1/G_{c} - 1/G ) \ ,
\ee
whereas weak coupling $G \le G_{c}$ corresponds to a symmetric 
phase ($\Phi_{0}=0$). 

\section{Dimensional reduction from the symmetric phase}

We compactify $x_{3}$ with periodicity $\beta$. Summing up the
momenta $k_{3}$ in the gap equation (\ref{3dgap}) leads to
\be
\frac{1}{G} = \frac{1}{(2\pi )^{2}} \int d^{2}k \, \frac{{\rm coth}
(\beta \sqrt{\vec k^{2} + \phi_{0}^{2}}/2)}{\sqrt{\vec k^{2} 
+ \phi_{0}^{2}}} \ .
\ee
The remaining 2-d integral involves again the cut-off $\Lambda_{2}$.
Consistency now requires the cut-off matching 
$\Lambda_{2} = 1/G_{c} = \Lambda_{3} / \pi^{2}$, and
$1/g = \beta (1/G - 1/G_{c})$. Thus we arrive at the 2-d fermion mass 
\be
m = \frac{2}{\beta} e^{- \pi \beta (1/G - 1/G_{c})}
= \frac{2}{\beta} e^{-\pi /g} \ ,
\ee
where $2/\beta$ takes the r\^{o}le of an effective cut-off in the 
reduced system. The 2-d correlation length $\xi = 1/m$ grows exponentially
as $\beta \to \infty$, hence we obtain dimensional reduction at {\em large}
$\beta$.

This result seems fine, but the procedure is not satisfactory
in view of our motivation: 
e.g.\ a non-perturbative treatment
at finite $N$ (on the lattice) should not start from the symmetric phase,
because this just shifts the problem of fine tuning to $d=3$.
Therefore we now focus on
{\em dimensional reduction from the broken phase}.

\section{A single domain wall}

Starting from the 3-d broken phase, the limit
$^{\lim}_{\beta \to 0} \beta /\xi = 2 \ln (1 + \sqrt{2})$
does not provide light fermions.  Hence we proceed differently
and generate a light 2-d fermion as the $k_{3}=0$-mode on a domain wall.
For the latter we make the ansatz $\Phi (x_{3}) = 
\Phi_{0} \tgh (\Phi_{0} x_{3})$, which is inspired by Refs.\ \cite{DHN}.
We choose $x_{2}$ as the time direction, hence the Hamiltonian reads
\be
H = \gamma_{2} [ \gamma_{1} \partial_{1} + \gamma_{3} \partial_{3} 
- \Phi (x_{3}) ] \ .
\ee
The ansatz $\Psi(x_{3}) e^{ik_{1}x_{1}} e^{-iEt}$
(and the chiral representation for $\gamma_{i}$) 
reveals one localized eigenstate of $H$,
\vspace*{-2mm}
\be
\Psi_{0}(x_{3}) = \sqrt{\frac{\Phi_{0}}{2}} \left(
\begin{array}{c}
0 \\ \cosh^{-1}(\Phi_{0}x_{3}) \end{array} \right)
\ee
with energy $E_{0} = -k_{1} >0$, i.e. a left-mover.
(On an anti-wall $-\Phi (x_{3})$
one would obtain a right-mover with $E_{0}=k_{1}>0$
and exchanged components in $\Psi_{0}(x_{3})$).

In addition there are bulk states (not localized in $x_{3}$),
\vspace*{-2mm}
\bd
\Psi_{k_{3}}(x_{3}) =
\frac{e^{ik_{3}x_{3}}}{\sqrt{2E(E+k_{1})}} \left( \!\!
\begin{array}{c}
i (E + k_{1}) \\ 
\Phi_{0} \tgh (\Phi_{0} x_{3}) -i k_{3}
\end{array} \!\! \right)
\ed
with $E = \pm \sqrt{\vec k^{2} + \Phi_{0}^{2}}$,
which form together with $\Psi_{0}$ an orthonormal 
basis for the 1-particle Hilbert space.

To verify the consistency of the wall profile we have to consider
the chiral condensate $\bar \Psi \Psi$. $\Psi_{0}$ does not contribute to it,
and if we sum up the bulk contributions of $E<0$ we reproduce exactly
$\Phi (x_{3})$, which confirms the self-consistency of this single brane
world.

In addition we are free to fill up  some of the $\Psi_{0}$ states.
Those with $E_{0} < \Phi_{0}$ are confined
to the $(1+1)$-d world, whereas states with $E_{0} \geq \Phi_{0}$ can escape
in the 3-direction. For the low energy observer on the brane this event
appears as a {\em fermion number violation}.

\section{A brane world with wall and anti-wall}

We now want to include both, $\Psi_{L}$ and $\Psi_{R}$, to be localized
on a wall and an anti-wall separated by $\beta$. For the 
profile we make the ansatz
\bea
\Phi (x_{3}) &=& \Phi_{0} \{ a [ \tgh_{+} - \tgh_{-}] -1 \} \nn \\
\tgh_{\pm} & := & \tgh (a \Phi_{0} [x_{3} \pm \beta /2 ])\ , \
a \in [0,1] .  \label{profwaw}
\eea
The ansatz for a bound state  with the same form as for single walls,
\be
\Psi_{0}(x_{3}) = c \left(
\begin{array}{c} \alpha_{1} \cosh^{-1}(a \Phi_{0}[x_{3} - \beta /2]) 
\\ \alpha_{2} \cosh^{-1} (a \Phi_{0}[x_{3} + \beta /2])
\end{array} \right) , 
\label{ansatzwaw}
\ee
implies the condition $\tgh (a \Phi_{0} \beta ) = a$. 
Hence
the parameter $a$ controls the brane separation, such that $a\to 0$
and $a \to 1$ correspond to $\beta \to 0$ resp. $\beta \to \infty$.

Remarkably, the Dirac equation in this background still has an analytic
solution, which is given by the ansatz (\ref{ansatzwaw}) with
\bea
&& \hspace*{-5mm} c = \frac{1}{2} \sqrt{ \frac{a \Phi_{0}}
{E_{0} ( E_{0} + k_{1})} } \ , \ \
\alpha_{1} = -i (E_{0} + k_{1}) \ , \nn \\
&& \hspace*{-5mm} \alpha_{2} = m = \sqrt{1 - a^{2}} \Phi_{0} \ , \ \ 
E_{0} = \pm \sqrt{k_{1}^{2} + m^{2}} \ . \nn
\eea
This $\Psi_{0}(x_{3})$ represents a Dirac fermion with components $\Psi_{L}$,
$\Psi_{R}$ localized on the wall resp.\ the anti-wall.
For a fast motion to the left (right) we have $0 < E_{0} \simeq
-k_{1} \ (+k_{1})$, so that the lower (upper) component dominates.
The mass $m$ measures the extent of the $L,R$ mixing. The limit $a \to
0$ does not provide a light fermion, $m= \Phi_{0}$. However,
the opposite limit $a \to 1$ achieves this, since $L,R$
mixing is suppressed,
\be
m \simeq 2 \, \Phi_{0} \, e^{- \beta \Phi_{0}} \ .
\ee
As in the case of the symmetric phase,
{\em large} $\beta$ implies $\xi \gg \beta$ and therefore 
dimensional reduction. A low energy observer in $d=1+1$ now
perceives a point-like Dirac fermion composed of $L-$ and $R-$modes.
On the other hand, a high energy observer in $d=2+1$ refers to the scale
$\Phi_{0}$ (the 3-d fermion mass) and observes a Dirac fermion with 
strongly separated $L-$ and $R-$constituents.

Also the bulk states can be determined
analytically \cite{BGW}.
Summing up again their $E<0$ contributions to $\bar \Psi \Psi$
leads to
\be
\frac{G}{N} \int dk_{3} \bar \Psi_{k_{3}} \Psi_{k_{3}} \vert_{E<0}
= \Phi(x_{3}) + {\cal A} \ ,
\ee
i.e.\ the desired result up to a term ${\cal A}$ \cite{BGW},
which has to be canceled by $\Psi_{0}$.
This requires all the bound states
with energies $E_{0} \le 
\Phi_{0}$ to be filled, i.e.\ exactly up to the threshold energy for
the escape into the third dimension.
Hence this wall anti-wall brane world does contain naturally
light fermions, but it is completely packed with them, so that its
physics is blocked by Pauli's principle.


We also checked if the wall and anti-wall repel or attract
each other, which could lead to a disastrous end of this brane world.
However, it turns out that the brane tension energy per fermion
always amounts to $\Phi_{0}$, hence it does not depend on the brane
separation, so this toy world is stable.


Finally we studied the possibility of adding a fermion mass term
$M \bar \Psi \Psi$ to the Lagrangian, so that the $Z(2)$ symmetry is
also {\em explicitly} broken in $d=3$ (which is actually realistic for a 
lattice formulation at finite $N$). This lifts the degeneracy of the
minima of $V_{eff}(\Phi )$. If we still insert the profile
(\ref{profwaw}), the condition 
for $\bar \Psi \Psi$ requires the bound fermion states to be
filled even beyond $\Phi_{0}$, hence in this case there is no
stable configuration.

\section{Summary}

We studied the reduction $d=3 \to 2$ in the Gross-Neveu model at
large $N$. In $d=3$ it splits into a symmetric phase with massless 
fermions and a broken phase, while the 2-d model is always broken and
asymptotically free. Our question was if a {\em natural} setting in $d=3$
can reduce to $d=2$ with light fermions ($m \ll \Lambda_{2}$), and thus
solve the hierarchy puzzle all by itself ?

Such a reduction from the symmetric phase does occur at {\em large}
$\beta$, but this is not the ``natural'' starting point we were looking for.

Hence we start from the 3-d {\em broken} phase and build a brane world
to achieve the desired mechanism. A single domain wall with a
standard profile has a localized left-handed bound state. The bulk
states then establish full self-consistency.

By means of a wall anti-wall pair we can accommodate left- and right-handed
fermion components, localized on the wall resp.\ anti-wall.
If their separation $\beta$ is large, this appears to the $(1+1)$-d observer
indeed as a point-like, light Dirac fermion. This world is stable,
but consistency requires to fill the bound states up to the threshold
for an escape into the 3-direction.
Therefore the construction is basically successful, but unfortunately 
this world does not enjoy any flexibility for physical processes.

\end{document}